\documentclass[11pt, oneside]{article}   	% use "amsart" instead of "article" for AMSLaTeX format
\usepackage{geometry}                		% See geometry.pdf to learn the layout options. There are lots.
\usepackage{graphicx}				% Use pdf, png, jpg, or eps§ with pdflatex; use eps in DVI mode
\usepackage{amssymb}
\usepackage{natbib}
\usepackage{listings}
\title{Delayed Recognition; the Co-citation Perspective}
\author{The Author}
\author{Wenxi Zhao\qquad Dmitriy Korobskiy \qquad George Chacko\\Netelabs, NET ESolutions Corporation (an NTT DATA Company)\\ McLean, VA 22102}

\begin{document}
\maketitle
%\section{}
%\subsection{}

%%%%%%%%%%%%%%%%%%%%%%%%%%%%%%%%%%%%%%%%%%%%%%%%%%%%%%%%%%%%%%%%%%%%%%%%%%%%%%%%%%%%%%%%%%%%%%%%%%%%%%%%%%%%%%%%%%%%%%%%%%%%%%%%%%%%%%%%%%%%%%%%%%%%%%%%%%%
% This is just an example/guide for you to refer to when submitting manuscripts to Frontiers, it is not mandatory to use Frontiers .cls files nor frontiers.tex  %
% This will only generate the Manuscript, the final article will be typeset by Frontiers after acceptance.   
%                                              %
%                                                                                                                                                         %
% When submitting your files, remember to upload this *tex file, the pdf generated with it, the *bib file (if bibliography is not within the *tex) and all the figures.
%%%%%%%%%%%%%%%%%%%%%%%%%%%%%%%%%%%%%%%%%%%%%%%%%%%%%%%%%%%%%%%%%%%%%%%%%%%%%%%%%%%%%%%%%%%%%%%%%%%%%%%%%%%%%%%%%%%%%%%%%%%%%%%%%%%%%%%%%%%%%%%%%%%%%%%%%%%

%%% Version 3.4 Generated 2018/06/15 %%%
%%% You will need to have the following packages installed: datetime, fmtcount, etoolbox, fcprefix, which are normally inlcuded in WinEdt. %%%
%%% In http://www.ctan.org/ you can find the packages and how to install them, if necessary. %%%
%%%  NB logo1.jpg is required in the path in order to correctly compile front page header %%%

\maketitle

% ABSTRACT
\begin{abstract}
A Sleeping Beauty is a publication that is apparently unrecognized for some period of time before experiencing sudden recognition by citation. Various reasons, including resistance to new ideas, have been attributed to such delayed recognition. We examine this phenomenon in the special case of co-citations, which represent new ideas generated through the combination of existing ones. Using relatively stringent selection criteria derived from the work of others, we analyze a very large dataset of over 940 million unique co-cited article pairs, and identified 1,196 cases of delayed co-citations. We further classify these 1,196 cases with respect to amplitude, rate of citation, and disciplinary origin and discuss alternative approaches towards identifying such instances.

\end{abstract}

\section{Introduction}

The term `Sleeping Beauty' has been used to describe an article that is not well cited in the early years after its publication but experiences a sharp increase in the rate at which it is subsequently cited~\citep{Raan2004}. An implication is that the new concept presented in such an article is `ahead of its time' and resistance to it delays recognition. Other causes for resistance and delayed recognition have been postulated that include (i) information overload from the large amount of information available, (ii)  modest communication skills of authors, (iii) insufficient promotion of ideas, (iv) conflict with existing theory and experimental data, (v) the author's position in the social hierarchy of science, (vi)  multiple discovery, (vii) the management structures of scientific institutions, (viii), and the conservative nature of establishments~\citep{Barber1961,Merton1963,Cole1970,Garfield1970a,Garfield1980a}. The Sleeping Beauty phenomenon, and variants of it, have been studied and debated with some degree of agreement that a fraction of the scientific literature exhibits citation kinetics that suggest delayed but eventual recognition of new ideas~\citep{Glanzel2003,Glanzel2004,Raan2004,redner_2005,Braun_2010,Li2014,Ke2015,Li2016,Song2018,sugimoto_mostafa_2018,ye_bornmann_2018,Raan2019}. 

Various approaches have been used to identify Sleeping Beauties and variants of it. Depth of sleep, length of sleep, and awake intensity as variables~\citep{Raan2004}, the Gini coefficient to examine later years of citation history~\citep{li_2014}, a parameter-free beauty coefficient ~\citep{Ke2015}, positional measures~\citep{costas2010}, and the citation angle by \cite{ye_bornmann_2018}. While earlier studies examined small datasets, subsequent ones considered large samples of the literature, for example, 22 million publications in \cite{Ke2015}.
	
The research cited above has focused on single publications, however, new ideas also result from combining two previously independent ones. The recognition of such novelty through combination can be examined by co-citation analysis~\citep{MarshakovaShaikevich1973,Small1973,Uzzi2013,Boyack2014,Wang2017,Bradley2020}. Tracing co-citations, therefore, provides another lens with which to study delayed recognition.  In precedent is a somewhat related use of co-citation analysis by \cite{zong_2018,teixeira2017sleeping} who sought to identify the so-called `princes' that awaken Sleeping Beauties. 

The measurement of delayed recognition by co-citation has been briefly explored by ~\cite{devarakonda_2020} in a study of 33.6 million reference pairs. The authors used simplified criteria derived from prior Sleeping Beauty studies on single publications~\citep{Ke2015,Raan2004,Raan2019}, reported 24 co-cited pairs all in the 99th percentile of co-citation frequencies, and proposed the term \emph{delayed co-citations} for such cases. This initial exploration, albeit at scale, only considered reference pairs where each member of a pair was in the 99th percentile of highly cited articles in Scopus.
In this article, we extend the work delayed co-citation to a much larger dataset, approximately 940 million pairs of articles. We refine the criteria in \cite{devarakonda_2020} and identify co-cited article pairs that exhibit delayed recognition using modifications of 
the techniques of \cite{Raan2004,Raan2019} and  \cite{Ke2015}. We also ask whether delayed co-citations are derived from Sleeping Beauty publications.

\section{MATERIALS AND METHODS}

We have previously described a dataset of 33.6 million cited pairs each belonging to the top 1\% of cited articles in the Scopus bibliography~\citep[Figure~2]{devarakonda_2020}. In the present study, we include all co-cited pairs from references cited by articles published in Scopus in the 11 year period, 1985-1995, not only those drawn from the top 1\% of cited articles. We developed methods to manage the expected volume of data using a combination of SQL, Cypher, and Python. Our code for parsing and updating Scopus XML data, a PostgreSQL schema for Scopus data, SQL, Cypher, and Python scripts used in this study are freely available from a Github repository~\citep{Korobskiy2019}.

To assemble and analyze a working dataset, we first exported 95,524,693 publication records from Scopus (all citation types) as a citation graph consisting of an edgelist and a nodelist, imported these data into a graph database (Neo4j) treating publications as nodes and citations as edges. After creating indexes to improve performance, we selected all publications of citation type `article' published in the years 1985-1995 (inclusive of both) that had at least five cited references each. In counting references, we only considered references with complete Scopus records. Incomplete references and those with cryptic placeholder identifiers were removed from the dataset. We also filtered rare cases in the data where a publication cites itself, or if the publication date of a cited reference was missing or greater than the publication date of its citing article. Selection of publications with at least 5 references was performed after curating references.

After initial comparison of SQL vs Cypher, we chose, on the basis of simplicity and performance, to use Cypher queries in Neo4j to generate all pairwise $n\choose 2$ combinations of an article's cited references. We de-duplicated these pairs across all articles to assemble a dataset of $\sim$940 million pairs (940,357,633 pairs), roughly 28 times larger than the dataset in \cite{devarakonda_2020}. We then calculated the frequency of co-cited pairs by dividing the data and processing batches in parallel using Neo4j and the GNU Parallel utility.  After tuning experiments on a test set of 1 million pairs using a Neo4j 4.0 in a Centos 7.5 virtual machine with 128 Gb of RAM and 16 vCPUs in the Microsoft Azure environment, we set the batch size to 1,000 pairs and the degree of parallelization to 15 cores. Under these conditions, it took roughly 11 min to compute co-citation frequencies for a batch of 1,000 pairs. We divided these 940 million pairs into 9 subsets of around 100 million pairs each and processed them at the rate of approximately 19 hours per subset.  

In illustration, the simple Cypher query for calculating co-citation frequencies of pairs in Neo4j is shown below. The input to the query is a csv file containing two columns of article identifiers with each row representing a co-cited pair.  
\vspace{2 mm}
\lstset{language=Pascal, basicstyle=\footnotesize} 
\begin{lstlisting}
UNWIND $input_data AS row
MATCH (a:Publication{
node_id: row.cited_1})<--(p)-->(b:Publication{node_id:row.cited_2})
RETURN row.cited_1 AS cited_1, row.cited_2 AS cited_2, 
count(p) AS scopus_frequency;
\end{lstlisting}

Frequencies thus calculated, were loaded back into PostgreSQL. For kinetic analysis, we selected all pairs with a co-citation frequency $>=$ 100 and calculated the kinetics of citation accumulation from the first possible year of co-citation for each pair through the year 2018, again in Neo4j.  Finally, for continuity, we set zero as the frequency for all years between the first possible year of co-citation and the last co-cited year (2018), with  missing frequency counts. Minor differences between the data in  \cite{devarakonda_2020} are due to more current data in Scopus in our study, and computing kinetic data through 2018 in this study. We compared small samples between the two datasets and confirmed that these minor differences in co-citation frequencies could be bridged by including citations from publications in 2019 and later. 

After generating a dataset of 940 million pairs, we applied three relatively conservative conditions to identify co-cited pairs of interest: (i) a minimum peak (annual) co-citation frequency for a pair of at least 20 (ii) a minimum total co-citation frequency of at least 100 (iii) a requirement both members of a co-cited pair should be published no earlier than 1970. We then identified delayed co-citation cases by setting two more conditions: (i) a minimum sleeping duration of 10 years as measured from the first possible year of co-citation (the more recent publication year of the two articles), (ii) during this sleeping period of 10 years or more, the average co-citation frequency should be at most 1 with no more than 2 co-citations in any one year. 

We also calculated the slope between the co-citation frequency of the awakening year and the peak frequency and modified the Beauty Coefficient ~\citep{Ke2015,devarakonda_2020}, which was designed to measure kinetics in single publications, to be relevant to co-citations by treating the first possible year of co-citation equivalently to the year of publication for a single article~\citep{devarakonda_2020}. 

To identify, single Sleeping Beauty publications, we narrowed the criteria of \cite{Raan2019} to consider only a single sleeping period of 10 years or greater; depth of sleep (average citation rate during sleep) of at most 1; an awakening period of 5 years; and an average co-citation frequency during the awakening period (which is defined as awakening citation intensity by van Raan) of at least 5. We also calculated the Beauty Coefficient ~\citep{Ke2015} for all single publications for comparison.

\section{RESULTS AND DISCUSSION}

\begin{figure}[h!]
\begin{center}
\includegraphics[width=10cm]{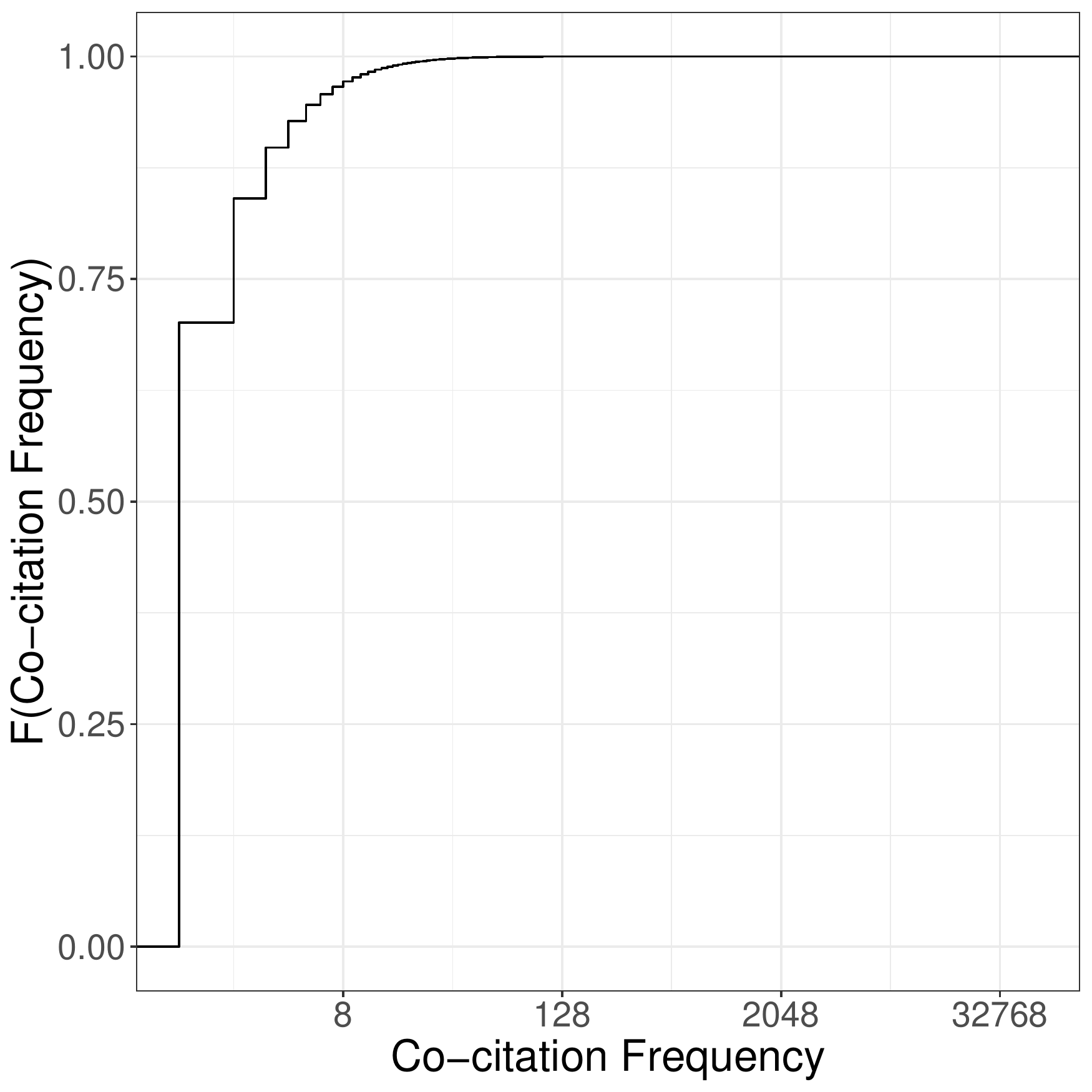}% This is a *.eps file
\end{center}
\caption{Frequencies of $\sim$940 million co-cited pairs drawn from Scopus 1985-1995. Pairwise combinations, $n\choose 2$, of references from articles indexed in Scopus (1985-1995), were generated as described in Materials and Methods. Total co-citation frequencies for these pairs, ranged from 1 to 52,471 with a median frequency of 1. The empirical cumulative distribution function (ECDF) was calculated from 940,357,633 co-citation frequencies and plotted against co-citation frequencies on a log\textsubscript{2} scale.}
\label{fig:fig1}
\end{figure}

In this study of delayed co-citations, we first examined cited references from 3,433,578 publications in the Scopus database. The criteria for selection of these publications were that they were classified as `article', that they were published in the period 1985-1995, and they contained at least 5 cited references each. We generated all possible co-cited pairs for the references in these articles and de-duplicated them across articles. since the same reference pair can occur in more than one article. Then we measured the co-citation frequency of each pair across the entire Scopus database by counting all co-citation events from the first possible year of co-citation onwards, Fig \ref{fig:fig1},Table \ref{tab:table1}).

\begin{table}[ht]
\caption{Distribution of ~940 million Co-citation Frequencies. The count of co-cited pairs in each frequency class as well as the percentage relative to the total number of 940,357,633 is shown. Counts include the lower bound in each class and exclude the upper bound. Add legend details}% title of Table
\centering % used for centering table
\begin{center}
\begin{tabular}{rll} 
\emph{f} Interval & Count & Percentage \\
\hline % inserts single horizontal line\
$<=$ 2 & 790,189,114 & 84.03 \\ 
2 -4 & 82,022,893 & 8.72 \\
4 -8 & 41,772,728 & 4.44 \\
8 -16 & 17,749,436 & 1.89 \\
16-32 & 6,429,234 & 0.68\\
32-64 & 1,704,908 & 0.18\\
64-128 & 385,923 & 0.041\\
128-256 & 81,164 & 0.0086\\ 
256-512 & 17,150 & 0.0018\\
512-1024 & 3,777 & 0.00040\\
1024-2048 & 948 & 0.00010\\ 
$> $ 2048 & 358 & 0.000038\\   
\hline 
\end{tabular}
\end{center}
\label{tab:table1} % is used to refer this table in the text
\end{table}

The data in Fig \ref{fig:fig1} show a highly skewed distribution of co-citation frequencies across a large dataset. Roughly 84\% of the pairs have a total co-citation frequency of 2 or less, and the 99th percentile is 16 although each pair had at least 10 years to accumulate co-citations. Even for a pair of articles from the most recent year in our data, 1995, this frequency of 16 corresponds to less than one co-citation per year on average. Thus, only a small fraction of pairs in these data have co-citation frequencies greater than 2 per year. One might consider that the reasons advanced for delayed recognition described in the Introduction could also contribute to such modest recognition or even acknowledgment of non-merit.

Beyond a high level understanding of the distribution of co-citation frequencies, however, we are interested in frequently co-cited publications, which are derived from highly cited publications~\citep{Small1973}, and 
are of interest to the community.  Thus, we subset the data using a conservative threshold of 100 for total co-citation frequency along with a peak annual co-citation frequency of at least 20. These criteria are analogous to those proposed by van Raan~\citep{Raan2004} and Redner~\citep{redner_2005}. After applying these two further restrictions, the number of co-cited pairs is reduced to 51,613 (approximately 0.055\% of the total number of pairs).

To find cases of delayed co-citation, we applied the following conditions to these 51,613 pairs: (i)  a co-cited paper should have slept for at least $10$ years and received no more than $2$ co-citations in each year during this sleeping period, which is defined as as the number of years from the first possible co-cited year to the first year that the pair receives more than $2$ co-citations. To be considered as a Sleeping Beauty, the awakening period that follows the sleeping period is characterized by (ii) a peak annual co-citation frequency of at least 20. These criteria when collectively applied, identified 1,196 cases of delayed co-citation, whose characteristics are summarized in Table \ref{tab:table2}. 

\begin{table}[ht]
\caption{Summary Statistics of 1,196 Delayed Co-citation Pairs. Criteria for selection were a minimum sleeping period of 10 years and a minimum peak of 20 citations in any year.}% title of Table
\centering % used for centering table
\begin{center}
\begin{tabular}{rllll} 
& Total Frequency & Sleep Duration & Slope & Beauty Coefficient* \\
\hline % inserts single horizontal line
Min &  20.00 & 10.00 & 0.21 & 34.21   \\ 
Q1  &  22.00 & 11.00  & 1.23 & 89.40   \\ 
Median & 26.00 & 14.00 & 1.7000 & 128.53   \\ 
Mean & 34.06 & 15.11 & 2.40 & 167.63   \\ 
3rd Qu & 36.00 & 17.00 & 2.67 & 190.93   \\ 
Max & 296.00 & 38.00  & 38.00  & 1678.62   \\ 
\hline
\end{tabular}
\end{center}
\label{tab:table2} % is used to refer this table in the text
\end{table}

Interestingly, these 1,196 pairs are derived from only 1,267 of a possible 2,392 individual publications indicating that some members of frequently co-cited pairs are found in multiple pairs. Indeed, we have previously noted that a pair of articles concerning methods in biochemistry, contribute to over 40,000 different co-cited pairs of frequency $>=$ 10~\citep{devarakonda_2020}.  

A logical question is whether any of these 1,267 individual publications would be classified as Sleeping Beauties. Applying van Raan's criteria (Materials and Methods), we identify 128 of these 1,267 publications as Sleeping Beauties. Interestingly, 27 of the 1,196 delayed co-citation pairs were cases where both members were Sleeping Beauties. Of these, the 1978 article by Rassias titled `On the stability of the linear mapping in Banach spaces' was a member of four different pairs. Thus, delayed recognition can occur without a requirement that at least one member of a co-cited pair with delayed recognition should have Sleeping Beauty characteristics. These  observations also suggest that while high-referencing fields such as biology~\citep{Small1980} might be advantaged by our selection criteria, the thresholds we set do not entirely exclude other fields. Accordingly, continuing this work with field normalization of co-citation frequencies, to the extent possible, is warranted.

In contrast to co-citation frequencies for delayed co-citations (Fig. \ref{tab:table2}), which range from 20-260; citation counts for the 1,267 publications that contribute to these 1,196 delayed co-citations range from 121 to 190,832 with 72 of these publications having citation counts of greater than 10,000. 

However, other co-citation frequencies do exceed the seemingly modest frequencies noted for delayed co-citations. For example, \cite{becke_densityfunctional_1993} and \cite{lee_development_1988}, a pair of articles from the field of physical chemistry, have been co-cited over 51,000 times but do not exhibit delayed citation kinetics. It should also be noted that these articles have individually been cited over 70,000 times each. Similarly, 1,357 pairs from the data shown in Fig \ref{fig:fig1} have co-citation frequencies greater than 1,000. 

We observe (Fig \ref{fig:fig1}), that the 90th, 95th, and 99th percentiles of co-citation frequencies in our dataset are 4, 6, and 16 respectively. In comparison. the 90th, 95th, and 99th percentile of citation frequencies of $\sim$10.7 million publications of type `article' in Scopus, published in the years 1970-1995, are 58, 96, and 254 respectively (roughly ten fold greater). What emerges is that delayed co-citations tend to have frequency profiles that are lower than those of other co-cited pairs, and single publications. This is not unexpected since co-cited frequencies cannot exceed the citation frequencies of the publications in these pairs but it does suggest that seemingly low co-citation frequencies should not be overlooked. 

\begin{figure}[h!]
\begin{center}
\includegraphics[width=10cm]{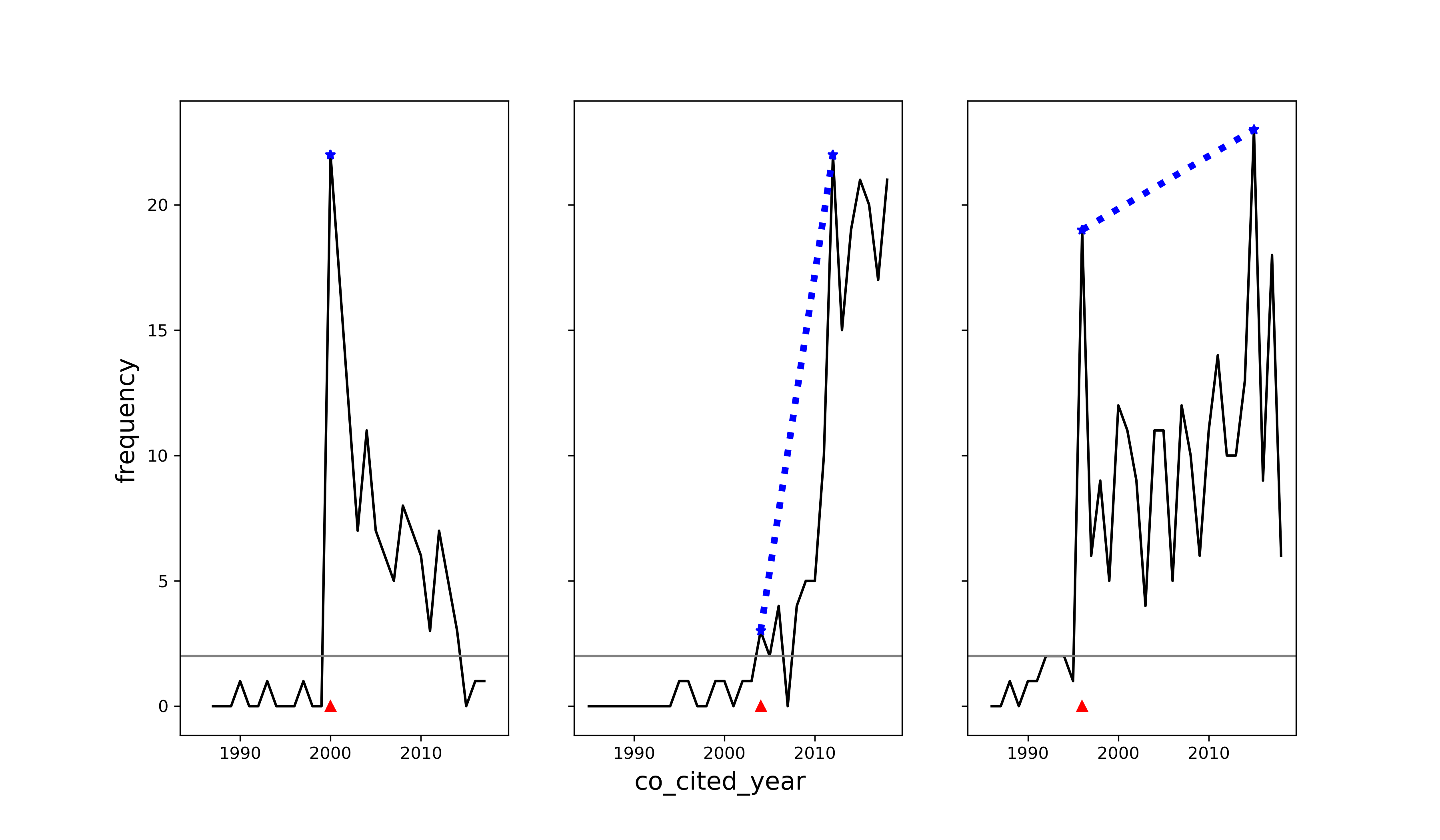}% This is a *.eps file
\end{center}
\caption{Kinetics of Co-citation Frequencies for Delayed Co-citations. Three sample plots are shown from 1,196 delayed co-citations selected for maximum slope (left panel). mean slope(middle panel), and minimum slope(right panel) of a line connecting the co-citation frequency of the awakening year to the co-citation frequency of the peak year . Total co-citation frequencies for these three plots were 131, 174, and 254, with peaks of 22, 22, and 23, and slopes of NA, 2.38, and 0.21 respectively. The red triangle marks the awakening year and the dotted blue line, the slope. The slope in the left panel is NA since the peak year is the awakening year. The article pairs shown above are (i) Spacetime as a membrane in higher dimensions (Gibbons 1987) \& An exotic class of Kaluza-Klein models (Visser 1985), (ii) Formulation of the reaction coordinate (Fukui 1070) \& Ab initio effective core potentials for molecular calculations. Potentials for main group elements Na to Bi (Wadt \& Hay 1985), (iii) A proposed grading system for arteriovenous malformations (Spetzler 1986) \& Arteriovenous malformations of the brain: Natural history in unoperated patients (Crawford et al. 1986).}
\label{fig:fig2}
\end{figure}

To examine rates of awakening, we also calculated the slope between the co-citation frequency in the first awakening year and the frequency of the peak year and noted a fairly broad range of slopes with a mean of 2.4 (Table \ref{tab:table2}). The kinetics of co-citation are visualized in Fig \ref{fig:fig2}, for three examples with the maximum slope, the mean slope, and the minimum slope observed.  

Of 1,196 delayed co-citations, the slope could not be computed for 10 pairs because the peak year was the year of awakening. This small number of cases, suggest sudden recognition of the concepts represented by these pairs (Table \ref{tab:table3}. These 10 pairs span the areas of LED technology, cosmology, immunology, psychology, and computational science. One publication from 1985 titled, ``An exotic class of Kaluza-Klein models" appears in 3 of 10 pairs and the author himself refers, in 1999, to `renewed interest due to the explosion of activity in the non compact extra dimensions variant of the Kaluza Klein model'~\citep{visser_1999}.

\begin{table}[ht]
\caption{Co-cited pairs with peak frequency in the the first year of awakening}% title of Table
\centering % used for centering table
\begin{center}
\begin{tabular}{ll} 
\hline % inserts single horizontal line
Years & Title \\
\hline 
\tiny{1974} & \tiny{Fundamental energy gap of GaN from photoluminescence excitation spectra} \\
\tiny{1971} &  \tiny{Absorption, reflectance, and luminescence of GaN epitaxial layers} \\
\hline
\tiny{1986} & \tiny{Dimensional reduction caused by a cosmological constant} \\
\tiny{1985} & \tiny{An exotic class of Kaluza-Klein models} \\
\hline
\tiny{1987} & \tiny{Spacetime as a membrane in higher dimensions} \\
\tiny{1985} & \tiny{An exotic class of Kaluza-Klein models} \\
\hline 
\tiny{1985} & \tiny{An exotic class of Kaluza-Klein models} \\
\tiny{1985} & \tiny{Do we live inside a domain wall?} \\
\hline
\tiny{1971} & \tiny{Mental rotation of three-dimensional objects} \\
\tiny{1976} & \tiny{Demonstration of a mental analog of an external rotation} \\
\hline
\tiny{1974} & \tiny{Biologic and clinical significance of cryoglobulins. A report of 86 cases} \\
\tiny{1980} & \tiny{Mixed cryoglobulinemia: Clinical aspects and long-term follow-up of 40 patients} \\
\hline
\tiny{1977} & \tiny{Imitation of Facial and Manual Gestures by Human Neonates} \\
\tiny{1979} & \tiny{Matching behavior in the young infant.} \\ 
\hline
\tiny{1978} & \tiny{Cognitive determinants of fixation location during picture viewing} \\
\tiny{1979} & \tiny{Framing pictures: The role of knowledge in automatized encoding and memory for gist} \\
\hline
\tiny{1983} & \tiny{Parst: A system of fortran routines for calculating molecular structure parameters (truncated)} \\
\tiny{1983} & \tiny{On enantiomorph‐polarity estimation} \\
\hline
\tiny{1980} & \tiny{Toward a positive theory of consumer choice} \\
\tiny{1973} & \tiny{On the psychology of prediction} \\
\hline 
\end{tabular}
\end{center}
\label{tab:table3} % is used to refer this table in the text
\end{table}

We also examined lesser co-citation frequencies, between 20 and 100, and observed 5,928,815 pairs. After removing pairs with (i) less than 10 years of kinetic data (the difference between publication year and peak year is less than 10 years) (ii) a negative Beauty Coefficient, which describes articles whose citations growing linearly with time or with a citation trajectory that is a concave function of time, (iii) without at least one peak of frequency 20, then the number reduced to 13,057 pairs. Of these 12,920 had only a single peak of 20 or greater that may be similar to `flash in the pan' citations~\citep{Li2013CitationCO,ye_bornmann_2018}. Given our focus on frequently co-cited pairs, we did not study these further.

An appealing alternative approach for delayed co-citations and Sleeping Beauties is the Beauty Coefficient. We have previously modified~\citep{devarakonda_2020}  the Beauty Coefficient~\citep{Ke2015} designed to measure kinetics in single publications, to be useful to the case of co-cited pairs. We computed the Beauty Coefficient for these 1,196 pairs observing a range of 34.21-1678.62. These data are summarized in Table \ref{tab:table2}.  Given co-citation frequencies being generally lower than citation frequencies, the top 15 Beauty Coefficient values of the 1,196 delayed co-citations range from 712.47-1678.62, which appear comparable to the top 15 described by Ke, all above 2,000.

Ke and colleagues comment that parameterized approaches in preceding studies have suffered from being somewhat arbitrary. The comment is fair, but arbitrariness may not have impeded discovery, for example Redner's work on the physics literature~\citep{redner_2005} with its selection threshold of 250 citations. Further, while the Beauty Coefficient is parameter free, the choice of selection threshold is left to the user leaving the door open for arbitrary selection thresholds. We consider this a strength of the measure since it can be used in contextual studies. The approach of van Raan is also intuitive and flexible but does not consider the maximum number of citations received as an important parameter to be tuned.  The cases with a sleeping period of ten years, and a citation rate of 5 for the next 5 years, would satisfy requirements for a Sleeping Beauty but are perhaps less noteworthy.

\begin{figure}[h!]
\begin{center}
\includegraphics[width=10cm]{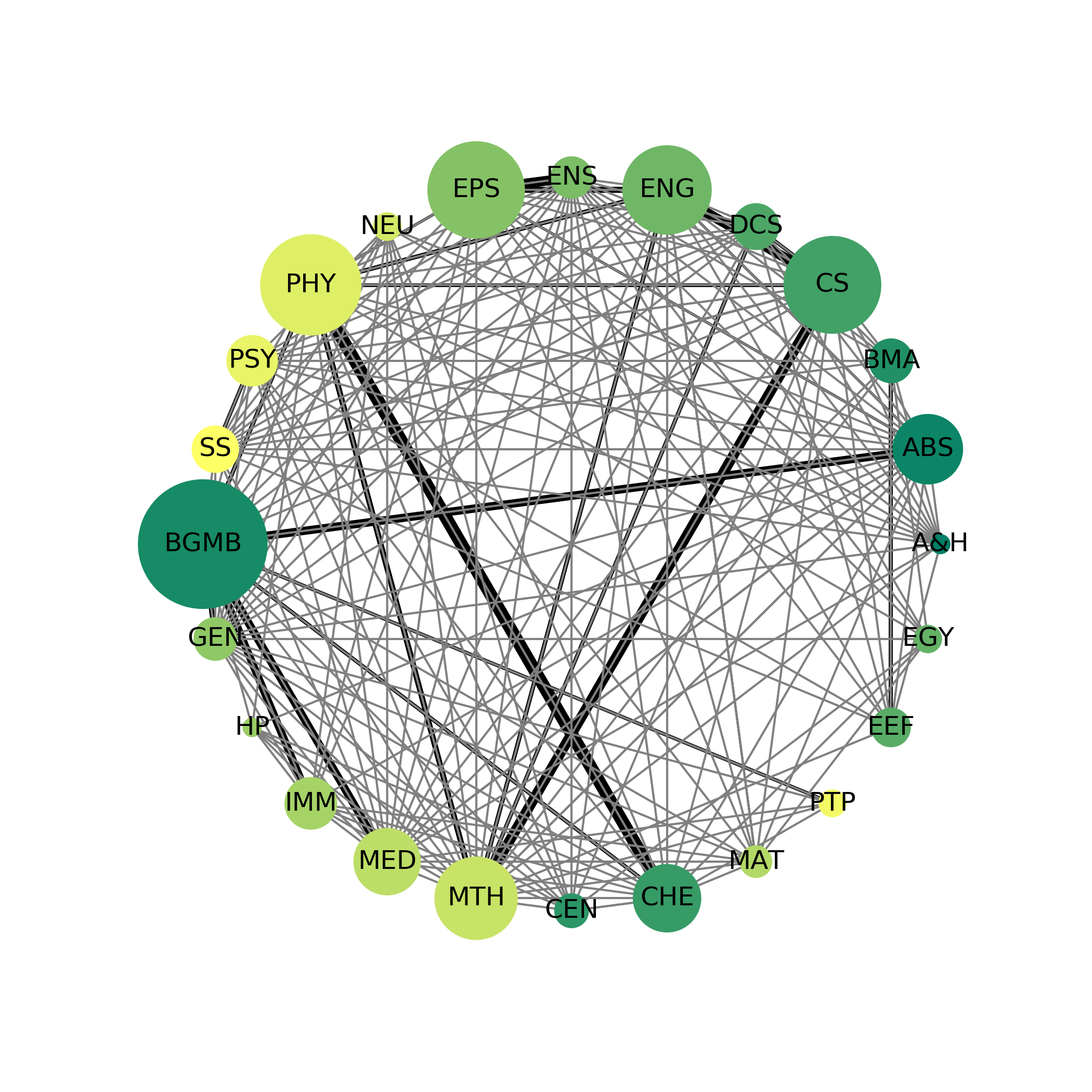}% This is a *.eps file
\end{center}
\caption{Disciplinary composition of 1,196 Delayed Co-citations. Each node represents a major subject area in the Scopus ASJC classification. Node size is scaled to the number articles in a given subject area. Edge thickness indicates the number of pairs that have one member in one each of the two nodes connected by the edge. Major subject areas are abbreviated in the graphic:
\emph{MTH} (Mathematics);
\emph{IMM} (Immunology and Microbiology);
\emph{HP} (Health Professions);
\emph{GEN} (General);
\emph{ENS} (Environmental Science);
\emph{ENG} (Engineering);
\emph{EPS} (Earth \& Planetary Sciences);
\emph{DCS} (Decision Sciences);
\emph{MAT} (Material Sciences);
\emph{CEN} (Chemical Engineering);
\emph{PSY} (Psychology);
\emph{PHY} (Physics and Astronomy);
\emph{NEU} (Neuroscience);
\emph{CS}( Computer Science);
\emph{A\&H} (Arts and Humanities);
\emph{SS} (Social Sciences);
\emph{MED} (Medicine);
\emph{EGY} (Energy);
\emph{CHE} (Chemistry);
\emph{ABS} (Agricultural \& Biological Sciences); 
\emph{BGMB} (Biochemistry, Genetics \& Molecular Biology);
\emph{BMA} (Business, Management, and Accounting);
\emph{EEF} (Economics, Econometrics and Finance)
\emph{PTP} (Pharmacology, Toxicology \& Pharmaceutics)}
\label{fig:fig3}
\end{figure}

Finally, to ask which fields these 1,196 delayed co-citations are found in, we mapped them to the All Science Journal Classification (ASJC) maintained by Scopus, which consists of 27 major subject area categories.  The data are represented in Figure \ref{fig:fig3} but should be interpreted in the light of these subject area labels being derived from journals and that an article may have more than one label. Even so, the data suggest that delayed co-citations, as we define them in our dataset are largely drawn from the domain of biochemistry, genetics, and molecular biology followed by physics, computer science, chemistry, and engineering. 

\clearpage
 
\section{CONCLUSION} In a large-scale exploration of the kinetics of co-citation (more than 940 million unique article pairs), we have identified 1,196 cases of delayed co-citation using criteria largely derived from the work of van Raan and Ke. We acknowledge that our selection criteria, while guided by positional statistics and intuitive preference, suffers from some degree of arbitrariness.  With all bibliometric data, coverage and data quality also influence discovery. Thus, we have tried to identify co-cited pairs of higher frequency since the trends in such cases are more likely to be reproducible across other data sources. Relaxing these conditions, will identify additional cases. Our goal was to identify a set of delayed co-cited pairs that can be studied, in the longer term, to understand the reasons for the patterns of citation. This future task will require a greater understanding of the fields in which such delayed co-citations occurred and ideally should be coupled to qualitative techniques. 
%\section{Additional Requirements}

\section*{Conflict of Interest Statement}
Data used in this study derive from the ERNIE project, which involves a collaboration with Elsevier. The content of this publication is solely the responsibility of the authors and does not necessarily represent the official views of the National Institutes of Health or Elsevier.  Elsevier staff did not have a role in design, manuscript-writing, or review and interpretation of results.

\section*{Author Contributions}

Wenxi Zhao: Conceptualization; Methodology; Investigation; Writing—Review and Editing. Dmitriy Korobskiy: Methodology; Writing – Review and Editing; George Chacko: Conceptualization; Methodology; Investigation; Writing—Original Draft; Writing—Review and Editing; Funding Acquisition, Resources; Supervision.

\section*{Funding} Research and development reported in this publication was partially supported by federal funds from the National Institute on Drug Abuse (NIDA), National Institutes of Health, U.S. Department of Health and Human Services, under Contract Nos. HHSN271201700053C (N43DA-17-1216) and HHSN271201800040C (N44DA-18-1216).

\section*{Acknowledgments}
We thank our Elsevier colleagues for their support of the ERNIE project. 

\section*{Data Availability Statement}
Data used in this study are restricted by a license from Elsevier Inc. Interested persons with a license for these data can use the code on our Github repository~\citep{Korobskiy2019} to reproduce our findings. 

\bibliographystyle{apa-good} 
\bibliography{sb_arxiv}
\end{document}